\documentclass[12pt]{article}
\usepackage[left=2.4cm,right=2.4cm,top=2.8cm,bottom=2.8cm]{geometry}
\usepackage[colorlinks=true,linkcolor=black,citecolor=black,urlcolor=blue,filecolor=black]{hyperref}

\usepackage[utf8]{inputenc}
\usepackage{amsmath}
\usepackage{amsfonts}
\usepackage{amssymb}
\usepackage[nosort]{cite}
\usepackage[usenames,table]{xcolor}
\usepackage{dsfont}
\usepackage{mathrsfs}

\usepackage{verbatim}

\begin{document}
\vspace{30pt}

\begin{center}


{\Large{\sc Consistent couplings between a massive spin-3/2 field and a partially 
massless spin-2 field}\\[10pt]}

\vspace{-5pt}
\par\noindent\rule{350pt}{0.4pt}


\vspace{20pt}

{\sc 
Nicolas Boulanger, Guillaume Lhost and Sylvain Thom\'ee\footnote{Research Fellow of the F.R.S.-FNRS (Belgium).}
}

\vspace{35pt}
{\it\small
Service de Physique de l'Univers, Champs et Gravitation,\\
Universit{\'e} de Mons -- UMONS,
20 place du Parc, 7000 Mons, Belgium}
\vspace{10pt}

{\tt\small 
\href{mailto:nicolas.boulanger@umons.ac.be}{nicolas.boulanger@umons.ac.be},
\href{mailto:guillaume.lhost@umons.ac.be}{guillaume.lhost@umons.ac.be},
\href{mailto:sylvain.thomee@umons.ac.be}{sylvain.thomee@umons.ac.be}
}

\vspace{40pt} {\sc\large Abstract} \end{center}

\noindent

We revisit the problem of constructing consistent interactions 
 between a massive spin-3/2 field and a partially massless graviton in 
 four-dimensional (A)dS spacetime. 
 We use the Stueckelberg formulation of the action principle for these fields 
 and find two non-trivial cubic vertices with less than two 
 derivatives, when going to the unitary gauge.
 One of the vertices is reminiscent of the minimal coupling of the massive 
 spin-3/2 field to gravity, except that now the graviton is partially massless.
\newpage


\section{Introduction}

The use of the Stueckelberg formulation for the problem of 
constructing consistent interactions between massive fields 
has proved very efficient, mainly through the works of Zinoviev 
and collaborators, see e.g. 
\cite{Zinoviev:2006im, Khabarov:2021djm, Zinoviev:2023vvz}
and references therein.
Some years ago in \cite{Zinoviev:2018eok}, Zinoviev constructed 
cubic couplings between a massive spin-3/2 field and a massive 
spin-2 graviton around (A)dS$_4$ background, with 
the assumption that the couplings should not bring more than one derivative.
Then, the partially-massless limit for the graviton was considered, 
bringing the conclusion that no cubic vertex with one derivative 
survived in this limit.

We wish to revisit this question, this time taking the spin-2 
field as partially massless from the very beginning of the analysis. 
Indeed, the operations of introducing interactions and taking a partially 
massless limit do not commute, in general.
In fact, in this Letter we report about a coupling between a partially 
massless spin-2 field and a massive spin-3/2 field that seems to have gone 
unnoticed in previous investigations, as far as we could see. 

In order to build consistent vertex involving massive fields 
(in the present case, the massive spin-3/2 field),
we use the method proposed in \cite{Boulanger:2018dau} that 
combines the cohomological reformulation of the Noether method 
for gauge systems \cite{Barnich:1993vg,Henneaux:1997bm}
with the Stueckelberg formulation for massive fields.
The advantage of the method \cite{Boulanger:2018dau} 
is that it exploits
the gauge structure of massless theories to describe interactions 
for massive fields. It proved useful in showing that dRGT gravity 
(see e.g. \cite{deRham:2014zqa} for a review) can be recast in a frame 
where the Einstein-Hilbert structure disappears, leaving only 
a Born-Infeld-like theory, with vertices obtained by contraction 
of products of the manifestly gauge-invariant field strengths in the 
Stueckelberg formulation of the free massive theory. 
In particular, in \cite{Boulanger:2018dau} 
the full list of cubic vertices of massive dRGT gravity theory 
was recovered, showing the usefulness of the method.

In this Letter, we use the cohomological method of \cite{Boulanger:2018dau} 
for the search of consistent couplings between 
a massive spin-$3/2$ field and a partially massless (PM) spin-$2$ field, 
also called PM graviton. The spacetime backgrounds considered are the 
Anti-de Sitter (AdS) and the de Sitter (dS) geometries where the 
cosmological constant $\Lambda$ is negative and positive, respectively.
In such spacetimes, a PM graviton possesses four propagating 
degrees of freedom and is characterized by a well-suited mass directly 
related to the cosmological constant $\Lambda\,$, 
see \cite{Deser:1983tm,Higuchi:1986py,Deser:2001us} and references therein.
In its PM phase, the graviton therefore propagates four degrees of freedom, 
exactly like the massive spin-3/2 field that carries four degrees of freedom. 
It is therefore natural to ask whether it is possible to 
elaborate a consistent gauge theory in which a PM graviton $k_{\mu\nu}$ 
and a massive spin-$3/2$ field $\psi_\mu$ are involved.

We study the problem in both dS ($\Lambda > 0$) 
and AdS ($\Lambda < 0$) backgrounds at a stroke, 
through the use of a parameter $\sigma$ that takes the value $+1$ in AdS 
and $-1$ in dS.

\section{Main results}

The main results we report in this Letter consist in the construction of two 
vertices expressed in the Stueckelberg formulation for both the PM spin-2 
and the massive spin-3/2 fields. 
In the unitary gauge where the Stueckelberg fields are set to zero, 
the first vertex is proportional to 
\begin{equation}
    \ell^{(1)} = 2\,\nabla_{[\mu}k_{\nu]\rho} \,{\overline{\psi}}{}^\mu \,
    \gamma^\rho \,\psi^\nu\;,
\end{equation}
where the spinor field $\psi_\mu$ satisfies the Majorana reality 
condition 
and denotes the field for the massive spin-3/2
particle (the spinor indices are left implicit), 
the Lorentz-covariant
derivative for the background geometry\footnote{We use conventions whereby 
the Lorentz-covariant derivative satisfies 
$[\nabla_\mu,\nabla_\nu]V^\rho = -2\sigma \,\lambda^2 \,
\delta^{\rho}{}_{[\mu}\,V_{\nu]}\,$, where $\sigma =\pm 1\,$. 
In other terms, the cosmological constant is $\Lambda = - 3 \,\sigma \,\lambda^2\,$
in four dimensions, where $\sigma =-1$ corresponds to dS$_4$ and 
$\sigma = 1$ to AdS$_4\,$. On a Dirac spinor $\psi\,$, 
we have $[\nabla_{\mu}, \nabla_{\nu}] \psi 
= - \tfrac{1}{2} \,\sigma \,\lambda^2\, \gamma_{\mu \nu }\,\psi\,$.} 
is denoted by the symbol $\nabla_\mu\,$, and $k_{\mu\nu}$ represents the 
PM spin-2 field. 
The components of the (A)dS background metric will be denoted $\overline{g}_{\mu\nu}\,$.
As usual, the four Dirac matrices are denoted by 
$\gamma^a\,$, $a=0,1,2,3\,$, 
and $\gamma^\mu:= \overline{e}^\mu{}_a\,\gamma^a\,$ featuring 
the components of the background (A)dS vierbein.

The other coupling is more interesting. 
In the unitary gauge, it reads
\begin{align}
    \ell^{(2)} = k_{\mu\nu}\,\mathcal{T}^{\mu\nu}\;,
    \quad \mathcal{T}^{\mu\nu} =&\;  
    \omega \,\overline{\psi}_\rho ~\gamma^{\rho(\mu}\,\psi^{\nu)} 
    + \overline{\psi}_\sigma \,\gamma^{\rho\sigma(\mu}
    \,\nabla_\rho \psi^{\nu)}\;,
\quad \omega := \sqrt{m^2+\sigma\lambda^2}\,\;,
\end{align}
where the real parameter $m$ is the mass of the spin-3/2 
field in AdS, 
in the sense that the limit $m\rightarrow 0$ is the limit 
where the spin-3/2 field enjoys a gauge symmetry that 
removes the helicity $\pm 1/2$ degrees of freedom, leaving only the 
helicity $\pm 3/2$ degrees of freedom on-shell.
The tensor $\mathcal{T}^{\mu\nu}$ is traceless and divergenceless 
on-shell:
\begin{equation}
    \overline{g}_{\mu\nu}\,\mathcal{T}^{\mu\nu} \approx 0\quad ,\qquad
    \nabla_\mu \mathcal{T}^{\mu\nu} \approx 0\;,
\end{equation}
where a weak equality is an equality that holds on the solutions 
of the field equations for the free theory. 
The above vertex 
$\ell^{(2)}=k_{\mu\nu}\,\mathcal{T}^{\mu\nu}\,$ 
induces a deformation of the gauge transformations on the 
physical fields $(k_{\mu\nu},\psi_{\mu})\,$ in the unitary gauge, 
given by
\begin{align}
   \delta_1\psi_{\mu} = -\psi^\nu\,\nabla_{\mu}\nabla_{\nu}\xi
   + \sigma\,\lambda^2\,\psi_{\mu}\,\xi\;,
   \qquad \delta_1 k_{\mu\nu} = 0\;,
   \label{eq:deltaeff1}
\end{align}
where we recall that the free, quadratic action $S_0[k,\psi]$ 
is invariant under \cite{Deser:1983tm,Higuchi:1986py,Deser:2001us}
\begin{align}
   \delta_0 k_{\mu\nu} = \nabla_{\mu}\nabla_{\nu}\xi 
   - \sigma\,\lambda^2\,\overline{g}_{\mu\nu}\,\xi\;,
   \qquad \delta_0 \psi_{\mu} = 0\;.
   \label{eq:deltaeff0}
\end{align}
From the knowledge of the quadratic and 
cubic actions $S_0[k,\psi]$ and 
$S_1[k,\psi]=\int \!d^4x \sqrt{\bar{g}}\,\ell^{(2)}\,$
in the unitary gauge, 
one readily finds the consistency of the deformation reported above:
\begin{align}
    \delta_0S_1[k,\psi]+\delta_1S_0[k,\psi] = 0\;.
    \label{eq:gaugeinvaeff}
\end{align}
As far as the deformation $\ell^{(2)}$ is concerned, notice from 
\eqref{eq:deltaeff1} and \eqref{eq:deltaeff0} 
that the transformation of the massive spin-3/2 field  
can be written as
$\delta_1 \psi_\mu=-\,\psi^\nu \,\delta_0 k_{\mu \nu}\,$, 
from which it is tempting to view the contravariant spinor
$\psi^\mu$ as a gauge-invariant quantity, defining the covariant 
field as 
$\psi_\mu:= \psi^{\nu}(\overline{g}_{\mu\nu}-\kappa \,k_{\mu\nu}+{\cal O}(\kappa^2))\,$,
where $\kappa$ is the deformation parameter that we take with 
units of length, that defines the perturbative 
expansion $S[\phi]=S_0[\phi] + \kappa\,S_1[\phi]+{\cal O}(\kappa^2)\,$, 
$\delta \phi = \delta_0\phi + \kappa\,\delta_1\phi+{\cal O}(\kappa^2)\,$
such that $\delta S[\phi]=0+{\cal O}(\kappa^2)\,$.
In this sense, like in Riemannian geometry, it would appear that 
a metric $g_{\mu\nu}:= \overline{g}_{\mu\nu}-\kappa \,k_{\mu\nu}+{\cal O}(\kappa^2)$ 
could be defined in terms of the (A)dS background metric and the PM spin-2 
field $k_{\mu\nu}\,$.

In the following section, we present the two deformations reported above 
in their Stueckelberg form, and explain how one can reach the unitary 
gauge at first order in perturbation, thereby reproducing the main results 
presented above.

\section{Consistent couplings in the Stueckelberg formulation}

In this section we first spell out the free model, then exhibit the 
first order interactions we found 
and finally explain how one can reach the unitary gauge at first 
order in deformation. 

\subsection*{The free model}

We want to investigate the couplings between a massive spin-3/2 field
and a PM spin-2 field. Our starting point will be the Stueckelberg 
formulation for these models. 
The Stueckelberg action for a massive spin-3/2 field and a PM spin-2 field
in (A)dS$_4$ reads
\begin{equation}
    \begin{aligned}
        S_0[k_{\mu\nu},B_\mu,\psi_\mu,\chi] = \int d^4x \,\sqrt{-\bar{g}}~
         &\Big( -\frac{1}{2} \nabla_\rho k^{\mu \nu} \nabla^\rho k^{\mu \nu }
        +  \nabla_\rho  k^{\mu \nu} \nabla_\mu  k^{\rho}_{~\nu} 
        -  \nabla_\mu k \nabla_\nu k^{\mu \nu }
        \\
        & + \frac{1}{2} \nabla_\mu k \nabla^\mu k
        + \frac{\sigma}{4} F_{\mu \nu} F^{\mu \nu } 
        -2 \sigma {\lambda}^2  (k_{\mu \nu } k^{\mu \nu} - \frac{1}{4} k^2) 
        \\
        & + 2{\lambda}~ [k \nabla_\mu B^\mu - k^{\mu \nu } \nabla_\mu B_\nu ] 
        + 3\lambda^2\, B^\mu B_\mu \Big) 
        \\
        &+
     \Big( - \frac{1}{2}~ \overline{\psi}_\mu ~\gamma^{\mu \nu \rho}~ \nabla_\nu \psi_\rho 
     + \frac{\omega}{2}~ \overline{\psi}_\mu ~\gamma^{\mu \nu} ~\psi_\nu
     - \frac{3 \omega }{2}~ \overline{\chi} \chi 
     \\
     & - \frac{3}{4} ~ \overline{\chi} ~\gamma^{\mu}~ \nabla_\mu \chi
     - \frac{3 m}{2}\, \overline{\psi}_\mu ~\gamma^\mu~ \chi 
      \Big)\,,
      \label{action stueck chap3}
    \end{aligned}
\end{equation}
where $F_{\mu\nu} = 2 \,\nabla_{[\mu} B_{\nu]}\,$.
For the spin-$2$ sector, we use the conventions of 
\cite{spin2twisted}. 
The vector field $B_\mu$ and the Majorana spinor $\chi$ are, 
respectively, the Stueckelberg companions of $k_{\mu\nu}$ 
and $\psi_\mu\,$.
The action \eqref{action stueck chap3} is invariant under 
the following Abelian gauge transformations:
\begin{equation}\label{eq:freegaugetransf}
    \begin{aligned}
    \delta_0 k_{\mu \nu} &= 2 \,\nabla_{(\mu} \varepsilon_{\nu)} 
    + \lambda\, \bar{g}_{\mu\nu} \,\pi\;,\quad
    \delta_0 B_\mu = \nabla_\mu \pi + 2\sigma \lambda\, \varepsilon_\mu\;,
    \\
    \delta_0 \psi_\mu &= \nabla_\mu \theta + \frac{\omega}{2}\, \gamma_\mu\, \theta\;,
    \quad \delta_0 \chi =m\, \theta\;.
    \end{aligned}
\end{equation}
It is useful to introduce the gauge-invariant quantities
\begin{equation}
\begin{aligned}
    \mathbf{\Psi}_\mu &:=
     \psi_\mu - \tfrac{1}{m}\,\nabla_\mu \chi
     - \tfrac{\omega}{2m}\,\gamma_\mu\, \chi\;,
     \\
     K_{\mu\nu\rho} &:= 
     2\, \nabla_{[\mu} k_{\nu]\rho}
    +  2 \,\lambda\, \bar{g}_{\rho[\mu} B_{\nu]}
    -  \tfrac{\sigma}{2\lambda}\,\nabla_\rho F_{\mu \nu} 
      \;,
\end{aligned}
\end{equation}
in terms of which the free action \eqref{action stueck chap3} 
can be written in a manifestly gauge invariant way:
\begin{equation}
\label{eq:StueckFreeAction}
\begin{aligned}
      S_0[k_{\mu\nu},B_\mu,\psi_\mu,\chi] 
      =\int d^4x \,\sqrt{-\Bar{g}}\,L_0 = 
       \frac{1}{2}\,\int d^4x \, \sqrt{-\Bar{g}}\,\Big(&-
      \overline{\mathbf{\Psi}}_\mu \;\gamma^{\mu \nu \rho} 
      ~\nabla_\nu  \mathbf{\Psi}_\rho 
      + \omega \, \overline{\mathbf{\Psi}}_\mu\; 
      \gamma^{\mu \nu} \;  \mathbf{\Psi}_\nu  
    \\
    &-\tfrac{1}{2}\;K_{\mu\nu\rho} K^{\mu\nu\rho} 
    + K^\mu K_{\mu} \Big)\;,
\end{aligned}
\end{equation}
where $K^\mu = \bar{g}^{\alpha \beta} K^{\mu}{}_{\alpha\beta}\,$.
The equations of motion obtained by extremizing the above action 
with respect to the fields $k_{\mu\nu}$ and $\psi_\mu$ are, 
respectively, 
\begin{align}
    \nabla_\rho K^{\rho(\mu \nu)}- \overline{g}^{\mu \nu} \nabla_\rho K^\rho + \nabla^{(\mu} K^{\nu)} \approx\; &0 \; , \\
    \nabla_\nu \overline{\mathbf{\Psi}}_\rho \,\gamma^{\mu \nu \rho} 
    + \omega\, \overline{\mathbf{\Psi}}_\nu \,\gamma^{\mu \nu} \approx\; &0\;.
\end{align}
These equations are useful in checking the consistency of the 
vertex $\ell^{(2)}\,$.

\subsection*{Interactions to first order}

We now report our main findings obtained following the method proposed in 
\cite{Boulanger:2018dau} for constructing interactions of massive fields 
in the Stueckelberg formulation. For the sake of conciseness, in this Letter 
we refrain from reviewing this method and instead spell out the results that 
can be checked without referring to the formalism developed in \cite{Boulanger:2018dau}. 

\begin{itemize}

\item

In the Stueckelberg formulation, the deformation $L^{(1)}_1$ of the Lagrangian 
that corresponds to the first vertex $\ell^{(1)}$ presented in the Introduction, 
reads 
\begin{align}
{L}^{(1)}_1 &= \overline{\mathbf{\Psi}}_\sigma\,\gamma^{\rho}\,
{\mathbf{\Psi}}_\alpha\,K^{\sigma\alpha}{}_\rho\;,
\label{eq:BIvertex}
\end{align}
up to trivial field redefinitions. The coefficient in front of it is, 
at this stage, arbitrary.
\item 
In the Stueckelberg formulation, the deformation ${L}^{(2)}_1$ of the Lagrangian 
that corresponds to the second vertex $\ell^{(2)}$ presented in the Introduction reads 
\begin{equation}
\label{eq:othervertex}
\begin{aligned}
    {L}^{(2)}_1 =\;&\omega\,
        \overline{\mathbf{\Psi}}_\mu \;\gamma^{\mu\nu}\;
        \mathbf{\Psi}^\rho \,k_{\rho\nu}
        - \tfrac{\omega^2}{m}\, \overline{\mathbf{\Psi}}^\mu\,  
        \gamma^{\nu} \,\chi\, k_{\mu\nu} + \tfrac{\omega^2}{m}\, \overline{\mathbf{\Psi}}_\mu\,  
        \gamma^{\mu} \,\chi\, k\\
        & \;+ \overline{\mathbf{\Psi}}_\sigma\,  
        \gamma^{\sigma\nu\rho}\,\nabla_\rho\mathbf{\Psi}^\mu\,k_{\mu\nu} 
        - \tfrac{2\,\omega}{m}\, \nabla_{[\rho} \overline{\mathbf{\Psi}}_{\sigma]}\, \gamma^{\sigma\alpha}  \chi\, k^\rho{}_{\alpha}
        - \tfrac{\omega}{m}\, \nabla_\rho \overline{\mathbf{\Psi}}_\sigma\, \gamma^{\rho\sigma} \chi \, k \\
        &\;- \tfrac{\sigma\,\omega}{2 \lambda}\, \overline{\mathbf{\Psi}}_\sigma  \gamma^{\rho\sigma} \mathbf{\Psi}^\alpha F_{\rho\alpha}
        - \tfrac{\sigma}{2 \lambda} \,\overline{\mathbf{\Psi}}_\sigma \gamma^{\alpha\rho\sigma} \nabla_\alpha \mathbf{\Psi}^\mu  F_{\rho\mu} 
        \;.
\end{aligned}
\end{equation}

Contrarily to the free Stueckelberg theory where the flat limit is smooth, 
in the interacting case one cannot take the limit $\lambda\rightarrow 0\,$, 
as the vertex is non-analytical in the constant~$\lambda\,$.

\item The above vertex  induces a deformation of the gauge transformations given by
\begin{align}
    \delta_1\psi_{\mu} =& \;  
    - \omega\, k_{\mu\nu}\,\gamma^\nu \,\theta
    + \omega\,\gamma_\mu \,\psi^\nu \,\varepsilon_\nu 
    - \tfrac{\omega\,\lambda}{m} \,\gamma_\mu \,\chi \,\pi
    - \lambda\,\psi_\mu \, \pi \nonumber\\
  & - \tfrac{\omega}{m}\,\gamma^\nu\,\nabla_{\nu}\varepsilon_{\mu}\,\chi
    - \tfrac{\omega}{m}\,\gamma_\mu\,\nabla_{\nu}\chi\,\varepsilon^\nu
    + 2\,\nabla_\mu\psi_{\nu}\,\varepsilon^\nu
    - \tfrac{\lambda}{m}\,\chi\,\nabla_{\mu}\pi
    \nonumber \\
  & - \tfrac{2}{m}\,\nabla_{\mu}\nabla_{\nu}\chi\,\varepsilon^\nu \;,
    \label{eq:delta1psi} \\
    \delta_1\chi =& \; 
    2\,m\,\psi_{\mu}\,\varepsilon^\mu
    - 2\,\varepsilon^\mu\,\nabla_{\mu}\chi 
    -\lambda\,\chi\,\pi \;,
   \label{eq:delta1chi} \\
  \delta_1 k_{\mu\nu} =& \;0 \;, \quad \delta_1 B_{\mu} = 0 \;. 
  \label{eq:deltah1}
\end{align}
The corresponding gauge algebra is
\begin{align}
 [\delta_{\theta},\delta_{\varepsilon}]\varphi &=  
 \delta_{\tilde\theta}\varphi\;,\qquad 
 \tilde\theta = 
  2\,\omega \,\gamma^{\mu} \theta\, \varepsilon_\mu\;,\\
 [\delta_{\theta},\delta_{\pi}]\varphi &= 
 \delta_{\overline\theta}\varphi\;,\qquad 
 \overline\theta = -2\,\lambda\, \theta\, \pi\;.
\end{align}

The redefinition of the gauge parameters that trivializes the gauge algebra is 
\begin{equation}
    \theta \quad\rightarrow\quad \theta 
    - \kappa\,\frac{\omega}{m}\,\gamma^\mu \,\chi \,\varepsilon_\mu 
    + \kappa\,\frac{\lambda}{m} \,\chi\, \pi\;.
\end{equation}
\end{itemize}

In order to express our results in the unitary gauge, we first 
need to explain how to reach the unitary gauge, in perturbation. 

\subsection*{Reaching the unitary gauge at first order in deformation}

The starting point is a free theory $S_0[\varphi^i,\chi^I]$ 
with a spectrum of fields $(\varphi^i,\chi^I)$ such that
the latter are Stueckelberg companions of the former. 
In other words, the action $S_0=\int d^nx \,{\cal L}_0$ 
is invariant under the gauge transformations
\begin{align}
    \delta_0 \varphi^i &= {R}_0{}^i{}_I\,\varepsilon^I 
    + {R}_0{}^i{}_{\alpha}\,\epsilon^\alpha\;, 
    \label{eq:StuPhi0}\\
    \delta_0 \chi^I &= m_I\,\varepsilon^I + 
    {R}_0{}^I{}_\alpha\,\epsilon^\alpha \;,
    \label{eq:StuChi0}
\end{align}
where we use De Witt's condensed notation. The gauge invariance under 
the Stueckelberg gauge parameters $\varepsilon^I$ imply the Noether 
identities
\begin{align}
    \frac{\delta {\cal L}_0}{\delta \chi^I} 
    \equiv -\frac{1}{m_I}\,R_0^{+}{}^i{}_I \frac{\delta {\cal L}_0}{\delta \varphi^i}\;,
    \label{eq:Noeth0}
\end{align}
where the operator $R_0^{+}{}^i{}_I$ denotes the adjoint of $R_0{}^i{}_I\,$.

We assume we also have a consistent, first order deformation of the action and 
gauge transformations, i.e., a functional 
$S_1[\varphi^i,\chi^I]=\int d^nx\, {\cal L}_1$ and gauge transformation laws  
\begin{align}
     \delta_1 \varphi^i &= R_1{}^i{}_I(\varphi,\chi)\,\varepsilon^I 
    + R_1{}^i{}_{\alpha}(\varphi,\chi)\,\epsilon^\alpha\;, 
    \label{eq:StuPhi1}\\
    \delta_1 \chi^I &= R_1{}^I{}_J(\varphi,\chi)\,\varepsilon^J + 
    R_1{}^I{}_\alpha(\varphi,\chi)\,\epsilon^\alpha \;,
    \label{eq:StuChi1}
\end{align}
such that 
\begin{align}
    \delta_1 S_0[\varphi^i,\chi^I] + \delta_0 S_1[\varphi^i,\chi^I] = 0\;. 
    \label{eq:gaugeinvarorder1}
\end{align}
Upon expanding the latter equation using \eqref{eq:StuPhi0}-\eqref{eq:StuChi1}
gives the following Noether identity associated with the gauge parameters 
$\varepsilon^I\,$:
\begin{align}
    \frac{\delta {\cal L}_1}{\delta \chi^I} \equiv 
    -\frac{1}{m_I}\,\Big(
     R_0^{+}{}^i{}_I\,\frac{\delta {\cal L}_1}{\delta \varphi^i}
    + [R_1^{+}{}^i{}_I(\varphi,\chi)
    -\frac{1}{m_J}\,R_1^{+}{}^J{}_I(\varphi,\chi)R_0^{+}{}^i{}_J]
    \,\frac{\delta {\cal L}_0}{\delta \varphi^i}
    \Big)\;.
\end{align}
Inserting this expression for $\frac{\delta {\cal L}_1}{\delta \chi^I}$
in the equation \eqref{eq:gaugeinvarorder1} yields
\begin{align}
    0 = \int d^nx\,\epsilon^\alpha \,\Big[ 
    {\cal R}^+_0{}^i{}_\alpha \frac{\delta {\cal L}_1}{\delta \varphi^i} +
    {\cal R}^+_1{}^i{}_\alpha(\varphi,\chi) \,
    \frac{\delta {\cal L}_0}{\delta \varphi^i}\Big]\;,
    \label{eq:remaininggaugeinva}
\end{align}
where 
\begin{align}
    {\cal R}_0{}^i{}_\alpha &= R_0{}^i{}_\alpha - 
    \frac{1}{m_I}\,R_0{}^i{}_I\,R_0{}^I{}_\alpha\;,
    \label{eq:calR0i}
    \\
    {\cal R}_1{}^i{}_\alpha(\varphi,\chi) &= 
       R_1{}^i{}_\alpha(\varphi,\chi)
     - \frac{1}{m_I}\,R_1{}^i{}_I(\varphi,\chi)\,R_0{}^I{}_\alpha
     - \frac{1}{m_I}\,R_0{}^i{}_I\,R_1{}^I{}_\alpha(\varphi,\chi)
     \nonumber \\
     &\quad +\;\frac{1}{m_Im_J}\,R_0{}^i{}_J\,R_1{}^J{}_I(\varphi,\chi)
     R_0{}^I{}_\alpha \;.  
    \label{eq:calR0I}
\end{align}
Equation \eqref{eq:remaininggaugeinva} expresses 
the $\epsilon^\alpha$-gauge invariance 
of the action $S[\varphi^i,\chi^I]= S_0[\varphi^i,\chi^I] 
+ g \, S_1[\varphi^i,\chi^I]\,$, to first order in perturbation 
--- here $g$ denotes the coupling constant used in perturbation. 
This equation is valid for an arbitrary field configuration, 
in particular it is valid when we set the Stueckelberg fields 
$\chi^I$ to zero. 
Using the following obvious equality
\begin{align}
    \frac{\delta {\cal L}_1}{\delta \varphi^i}(\varphi,\chi=0) = 
    \frac{\delta \check{{\cal L}}_1}{\delta \varphi^i}(\varphi)\;,
    \qquad \check{{\cal L}}_1 := \left. {\cal L}_1 \right|_{\chi = 0}\;,
\end{align}
gives 
\begin{align}
    0 = \int d^nx\,\epsilon^\alpha \,&\Big[ 
    {\cal R}^+_0{}^i{}_\alpha \frac{\delta \check{{\cal L}}_1}{\delta \varphi^i} +
    \check{\cal R}^+_1{}^i{}_\alpha(\varphi) \,
    \frac{\delta \check{{\cal L}}_0}{\delta \varphi^i}\Big]\;,
    \label{eq:remaininggaugeinvacheck}
    \\
    & \check{{\cal L}}_0 := \left. {\cal L}_0 \right|_{\chi = 0}\;,\quad 
    \check{\cal R}_1{}^i{}_\alpha(\varphi) 
    := {\cal R}_1{}^i{}_\alpha(\varphi,\chi=0)\;.
    \nonumber
\end{align}
In its turn, the latter equation expresses the gauge invariance, 
to first order in perturbation, of the reduced action 
$\check{S}[\varphi^i] =  S_0[\varphi^i,\chi^I=0] 
+ g \, S_1[\varphi^i,\chi^I=0]$ under the gauge transformations
\begin{align}
    \delta_\epsilon \varphi^i = [{\cal R}_0{}^i{}_\alpha \,
                 + g\,
                 \check{\cal R}_1{}^i{}_\alpha(\varphi) ]\,\epsilon^\alpha
                 + {\cal O}(g^2)\;.
    \label{eq:effectivetransformationsInGeneral}
\end{align}

In the particular case studied in this Letter where we have 
the physical fields $\varphi^i=\{k_{\mu\nu},\psi_\mu\}$ and the Stueckelberg 
fields $\chi^I=\{B_{\mu},\chi\}\,$ with the gauge transformations 
at zeroth and first order given in \eqref{eq:freegaugetransf} and 
\eqref{eq:delta1psi}-\eqref{eq:deltah1}, respectively, 
we find the equation \eqref{eq:gaugeinvaeff}:  
the reduced action is invariant under \eqref{eq:deltaeff1}-\eqref{eq:deltaeff0} as announced in the Introduction, where we renamed the scalar parameter 
$\pi$ into $\xi\,$, absorbing in it the constant factor $-\frac{\sigma}{\lambda}\,$.

From the above-derived formula \eqref{eq:effectivetransformationsInGeneral} 
for the gauge transformations $\delta_\epsilon \varphi^i$ 
that leave invariant the reduced action, 
one can make an observation on the corresponding transformations 
of the Stueckelberg field strengths
\begin{align}
    \Phi^i := \varphi^i - \tfrac{1}{m_J} \,{R}_0{}^i{}_J \,\chi^J\;.
\end{align}
As is well-known, these quantities are invariant under the 
Stueckelberg transformations
\begin{align}
    \delta^{\varepsilon}_0 \varphi^i &= {R}_0{}^i{}_I\,\varepsilon^I\;,\qquad
    \delta^{\varepsilon}_0 \chi^I = m_I\,\varepsilon^I\;.
\end{align}
Under the complete transformation laws \eqref{eq:StuPhi0}, 
\eqref{eq:StuChi0}, \eqref{eq:StuPhi1} and \eqref{eq:StuChi1}, 
the Stueckelberg field strengths $\Phi^i$ transform as 
\begin{equation}
\begin{aligned}
    \delta \Phi^i =& \;\mathcal{R}_0{}^i{}_\alpha \epsilon^\alpha 
    + g\, \Bigl( R_1{}^i{}_\alpha(\varphi,\chi) \epsilon^\alpha
     +R_1{}^i{}_I(\varphi,\chi)  \varepsilon^I \\
     &- \frac{1}{m_I}\,R_0{}^i{}_I\,R_1{}^I{}_\alpha(\varphi,\chi) \epsilon^\alpha -\frac{1}{m_J}\,R_0{}^i{}_J\,R_1{}^J{}_I(\varphi,\chi)\varepsilon^I \Bigr)+ {\cal O}(g^2)\;.
     \label{eq:deltaPhii}
\end{aligned}
\end{equation}
If, on the right-hand-side of the above formula, 
one sets $\chi^I=0$ and 
$\varepsilon^I = -\frac{1}{m_I}{R}_0{}^I{}_\alpha \,\epsilon^\alpha$ 
for the residual $\varepsilon^I(\epsilon)$ parameters that preserve the unitary gauge 
$\chi^I=0$ at zeroth order in perturbation, 
it turns out that one exactly recovers the expression for the 
$\delta_\epsilon \varphi^i$ transformations \eqref{eq:effectivetransformationsInGeneral}. 

In other words, one could turn the argument around and 
get a heuristic way of producing the right-hand 
side of formula \eqref{eq:effectivetransformationsInGeneral}: 
by demanding that the operations of setting the fields $\chi^I$ 
to zero and performing gauge transformations commute on the 
Stueckelberg field strengths, i.e.,  imposing
$(\left.\delta \Phi^i)\right|_{\chi=0,\varepsilon=\varepsilon(\epsilon)}
=\delta_\epsilon (\left.\Phi^i\right|_{\chi=0})\,$.

\section{Conclusions and outlook}

The first vertex $L_1^{(1)}$ 
that we presented above in \eqref{eq:BIvertex}  
does not deform the Stueckelberg 
gauge transformations \eqref{eq:freegaugetransf}. It is exactly invariant 
under the latter transformations. The second vertex $L_1^{(2)}$
given in \eqref{eq:othervertex} is more interesting 
in the sense that it truly deforms the transformations
\eqref{eq:freegaugetransf}. The first term of on the right-hand side of 
\eqref{eq:delta1psi} is reminiscent of the local supersymmetry transformations 
in AdS background, giving the minimal 
deformation of the mass-like term on the right-hand side of  
$\delta_0 \psi_\mu = \nabla_\mu \theta + \frac{\omega}{2}\, \gamma_\mu\, \theta\,$, 
see \eqref{eq:freegaugetransf}. 
Correspondingly, the first term on the right-hand side of 
\eqref{eq:othervertex}
is the minimal deformation of the mass-like term 
for the spinor $\mathbf{\Psi}_\mu$ in the free action \eqref{eq:StueckFreeAction}.
However, contrarily to the situation in supergravity theories, 
there is no deformation proportional 
to the linearised ``spin-connection" $\nabla_{[\mu}k_{\nu]\rho}\,$
in \eqref{eq:othervertex} or in \eqref{eq:delta1psi}.

It will be interesting to investigate the consistent interactions 
among the fields of the enlarged spectra given in \cite{Garcia-Saenz:2018}.
There it was shown that the doublet $(k_{\mu\nu},\psi_\mu)$ 
consisting of a PM spin-2 and a massive spin-3/2 field studied in the present 
Letter must be completed with a masseless spin-$(3/2,1)$ doublet in order 
to carry the action of supersymmetry.
We hope to report soon on the interactions among these four fields. 
In the paper \cite{Buchbinder:2019olk}, on the other hand, it was found that
the partially massless doublet $(5/2, 2)$ can be completed 
with a massless doublet $(2, 3/2)$ in order to carry the action 
of supersymmetry. We intend to investigate the consistent 
couplings among those fields in a future work. 
\appendix

\section*{Acknowledgments}

It is a pleasure to thank Yurii Zinoviev for useful comments and 
discussions. The work of N.B. was partially supported by the FNRS grant 
No. T.0022.19 ``Fundamental issues in extended gravity".
The work of S.T. was partially supported by the FNRS ASP fellowship 
FC 54793 MassHighSpin. The work of G.L. was partially supported by 
the FNRS grant F.4503.20.

\providecommand{\href}[2]{#2}\begingroup\raggedright\endgroup

\end{document}